\documentclass[prb,preprint]{revtex4}

\usepackage{amsmath}
\usepackage{graphicx}
\usepackage{epstopdf}
\usepackage{upgreek}

\newcommand{\Bv}{\mathbf{B}}
\newcommand{\Jv}{\mathbf{J}}

\newcommand{\grad}{\boldsymbol{\nabla}}

\begin{document}

\title{Equation of the field lines of an axisymmetric multipole with a source surface}
\author{S. G. Gregory}
\email{sgregory@astro.caltech.edu}
\affiliation{Astrophysics Group, School of Physics, University of Exeter, Exeter, EX4 4QL, United Kingdom}
\affiliation{California Institute of Technology, MC 249-17, Pasadena, CA 91101, U.S.A.}


\begin{abstract}
Optical spectropolarimeters can be used to produce maps of the surface magnetic fields of
stars and hence to determine how stellar magnetic fields vary with stellar mass,
rotation rate, and evolutionary stage. In particular, we now can map the surface magnetic fields of 
forming solar-like stars, which are still contracting under gravity and are surrounded by a disk of gas 
and dust. Their large scale magnetic fields are almost dipolar on some stars, and there is evidence 
for many higher order multipole field components on other stars. The availability of new data has renewed 
interest in incorporating multipolar magnetic fields into models of stellar magnetospheres. I describe the 
basic properties of axial multipoles of arbitrary degree $\ell$ and derive the equation of the field lines in 
spherical coordinates. The spherical magnetic field components that describe the global stellar field topology 
are obtained analytically assuming that currents can be neglected in the region exterior to the star, and 
interior to some fixed spherical equipotential surface. The field components follow from the solution of 
Laplace's equation for the magnetostatic potential.
\end{abstract}

\maketitle

\section{Introduction}
The solution of Laplace's equation by separation of variables for the electrostatic potential in a region external to 
a charge distribution is a standard topic in graduate and undergraduate courses in electromagnetism.\cite{jea27,ble76,hea95,wyl99}
As an example application, the long-range interaction between molecular charge clouds can 
be determined via a multipole expansion of the electrostatic potential obtained by solving Laplace's 
equation.\cite{gra84} The solution for the magnetostatic potential in a region devoid of 
current sources is equivalent to the electrostatic case.\cite{gra84}

In this paper I show how this approach can be adapted to construct models of the large scale magnetic fields of stars and planets. 
I will derive the equation for the field lines of an axial magnetic multipole of arbitrary degree $\ell$ with a source surface. The source
surface is a spherical surface of radius $R_s$ used by the solar physics community to mimic the effects of the solar wind. The solar
wind is continuous streams of outflowing charged particles that open the large scale magnetosphere (the region external to a star, or 
equivalently a planet, consisting of closed magnetic field lines) of the Sun.\cite{alt69,sch69} At $R_s$ the field is assumed to be purely 
radial. This model has been successfully adapted to produce models of stellar magnetospheres via field extrapolation 
from observationally derived magnetic surface maps.\cite{jar02,gre06,gre08} As shown in Fig.~\ref{fig_colored_dipole} multipole magnetic 
fields with a source surface boundary condition incorporate regions of closed field line loops, as well as regions of open field lines along 
which outflows (stellar winds) are launched. A basic assumption of the model is that currents can be neglected in the region external to the 
star, and interior to $R_s$. With this assumption and with the source surface boundary condition, the complicated problem of solving 
Poisson's equation to determine the magnetic field components is bypassed. Instead, the $\Bv$-field is derived from the general solution 
of the familiar Laplace's equation.

The layout of the paper is as follows. In Sec.~\ref{intro} some recent results from the study of stellar magnetic fields are presented. I define 
the field components of a multipolar large scale stellar (or equivalently a planetary) magnetosphere in 
Sec.~\ref{comps}. In Sec.~\ref{comps_part2} I discuss how models of stellar magnetospheres can 
account for the distortion of the large scale magnetic field caused by stellar outflows (winds) by the use of the source surface boundary condition,
and demonstrate in Sec.~\ref{comps_part3} how the field components are modified. The modified field components are derived
from the solution of Laplace's equation in spherical coordinates, subject to the assumptions discussed at the beginning of Sec.~\ref{comps_part3}. 
In Sec.~\ref{equ} I solve the differential equation for the path of the field lines for an arbitrary multipole $l$ with a source surface. I summarize the main results 
in Sec.~\ref{sum}.

Throughout the paper I consider only stellar/planetary magnetic fields that are in a ``potential state.'' This terminology, which is common in the solar/stellar 
physics literature, refers to a magnetic field in which the current density $\Jv =\mathbf{0}$ everywhere within the stellar 
magnetosphere, and therefore the field $\Bv$ can be written in terms of the gradient of a magnetostatic scalar potential.

\section{\label{intro}Stellar magnetic fields}
Surface magnetic field maps derived using the technique of Zeeman-Doppler imaging\cite{don09} have been obtained for a 
large number of stars of differing age and mass. Zeeman-Doppler imaging is a tomographic technique that allows stellar magnetic maps to be 
constructed by monitoring distortions in spectral lines recorded in circularly polarized light. Large scale stellar magnetospheres show varying 
degrees of complexity, from the simple well ordered almost dipolar field of a low mass star\cite{don06science} (V374\,Peg, $\sim0.3\,{\rm M}_{\odot}$) to the highly 
complex multipolar field of a high mass star\cite{don06} ($\tau$ Sco, $\sim15\,{\rm M}_{\odot}$).\cite{sno94} In Fig.~\ref{fig_map} we show an example of a 
magnetic map derived using Zeeman-Doppler imaging of the forming star V2129~Oph ($\sim$1.35$\,{\rm M}_\odot$).\cite{don07} Forming stars 
are still contracting under gravity. At this early phase of stellar evolution, the temperature/pressure in their cores is not yet sufficient 
for the fusion of hydrogen into helium to begin. The large scale magnetic field of V2129\,Oph was found to be 
dominantly octupolar.\cite{don07} Also shown in Fig.~\ref{fig_map} 
is a numerical extrapolation showing the three-dimensional field topology constructed from the magnetic map (see Ref.~\onlinecite{gre10} for 
a review of the numerical technique). 

Motivated in part by the availability of observational data, which for the first time is allowing variations in stellar magnetic field 
topologies with stellar age, mass, and rotation period to be probed, new theoretical models that incorporate multipolar magnetic 
fields have been developed.\cite{gre06,gre08} The basic properties of multipolar magnetospheres have been explored for many 
years by geophysicists. In Ref.~\onlinecite{wil87} an expression for the field lines of an axisymmetric (axial) multipole of arbitrary degree 
$\ell$ is derived (see also Refs.~\onlinecite{jef88, bac88} and Ref.~\onlinecite{wil88} for an extension to the field lines of sectorial multipoles). 
In the context of this paper the order $\ell$ of a magnetic multipole can be thought of as the number of polarity changes in the surface field 
between the north and south pole of the star along a line of constant longitude (see Fig.~\ref{color_plots}). By ``equation of the field lines'' I 
mean an expression of the form $r=r(\theta)$ which describes the path (shape) of the field lines in a spherical coordinate system. The 
simple form $r=r(\theta)$ is valid for field lines within a particular meridional plane (planar field lines). In this paper I restrict our
attention to axial multipoles, which generate planar field lines. For 
example, an axial dipole ($\ell=1$) stellar magnetic field has spherical components 
\begin{align}
B_r &= B^{1,{\rm pole}}_{\ast}\left(\frac{R_\ast}{r}\right)^3\cos{\theta} \label{br}\\
B_\theta &= \frac{1}{2}B^{1,{\rm pole}}_{\ast}\left(\frac{R_\ast}{r}\right)^3\sin{\theta} \label{btheta}
\end{align}
where $R_\ast$ is the stellar radius and $B^{1,{\rm pole}}_{\ast}$ is the strength of the dipole at the rotation pole of the star.\cite{gre10} 
The field components are sketched in Fig.~\ref{fig0}. 

Throughout this paper $r$ and $\theta$ are standard spherical polar coordinates, as illustrated in Fig.~\ref{fig0}, with $\theta$ 
measured from the rotation pole of the star; $\theta=0$ corresponds to the stellar rotation pole, and $\theta=\pi/2$ corresponds to the equatorial plane. The 
origin of the coordinate system is at the center of the star, and thus the stellar surface corresponds to $r=R_{\ast}$. Only azimuthally symmetric
multipoles, known as axial multipoles, are considered in this paper, that is, $B_\phi=0$.

The differential equation describing the path of the field lines is$\cite{wil88,gre06}$
\begin{equation} 
\frac{B_r}{dr}=\frac{B_\theta}{rd\theta},
\label{sphere_fieldline}
\end{equation}
which, upon substituting for $B_r$ and $B_\theta$ from Eqs.~\eqref{br} and \eqref{btheta} and integrating, yields the well known result 
(see, for example, Ref.~\onlinecite{wil87})
\begin{equation}
r={\rm constant} \times \sin^2{\theta}.
\label{int_const}
\end{equation}

If a particular closed field line loop of the dipole reaches a 
maximum radial extent of $r_{\max}$ in the stellar equatorial plane, where $\theta=\pi/2$ (see Fig.~\ref{fig0}), then 
the integration constant in Eq.~(\ref{int_const}) is equal to $r_{\max}$. Thus for a dipole the equation of the particular closed field line is 
\begin{equation}
r=r_{\max}\sin^2{\theta}.
\label{also_electric}
\end{equation} 
Different values of $r_{\max}$ correspond to different field lines. Closed field lines with larger values of $r_{\max}$ have footpoints at 
higher latitude (that is, at smaller co-latitude $\theta$). As shown in Fig.~\ref{fig_colored_dipole} the field line footpoints are where the 
closed loop under consideration connects to the stellar surface. (Open field lines have only a single footpoint on the star.) 

Equation~(\ref{also_electric}) also describes the field lines of a point electric dipole, and 
Eq.~(\ref{sphere_fieldline}) can be written in other coordinates for both the magnetostatic\cite{kir86,kno00}
and electrostatic cases. The later expression is found in many electromagnetism textbooks such as Ref.~\onlinecite{smy50}, their 
equation (2) in their section 1.08. The 
equation of the field lines for an arbitrary order finite electric linear 
multipole, for example, can be derived from the differential equation $E_x/dx=E_y/dy=E_z/dz$.\cite{smy50,kri85,kir85} 

\section{The magnetic field components of axial multipoles}\label{comps}
The large scale magnetic field external to a star or a planet is generated due to dynamo action and the distribution 
of current sources within its interior. The region external to the star can be assumed to be source free, an approximation which is 
discussed in Sec.~\ref{comps_part2}. The spherical magnetic field components of stellar magnetospheres were derived in 
Ref.~\onlinecite{gre10} by an expansion of the magnetostatic scalar potential using this assumption and the 
work of Refs.~\onlinecite{gra79} and \onlinecite{gra84}. I do not reproduce the lengthly derivations here.

In this paper I consider axial multipoles where I have chosen the space-fixed $z$-axis of the star (the stellar rotation axis) as the 
symmetry axis of the multipole being considered (see Ref.~\onlinecite{gre10} for details). As shown in Fig.~\ref{fig0} the field lines 
of axial multipoles are reflectionally symmetric in the $x$-axis, which lies in the stellar equatorial plane, and rotationally symmetric 
about the $z$-axis. The spherical field components of an axial multipole ($B_{\phi}=0$) of order $\ell$ are
\begin{align}
B_r &= B^{\ell,{\rm pole}}_{\ast} \left(\frac{R_{\ast}}{r}\right)^{\ell+2} P_\ell(\cos{\theta}) \label{Br} \\
B_{\theta} &= \frac{B^{\ell,{\rm pole}}_{\ast}}{\ell+1}\left(\frac{R_{\ast}}{r}\right)^{\ell+2}P_{\ell1}(\cos{\theta}), \label{Btheta} 
\end{align}
where $B^{\ell,{\rm pole}}_{\ast}$ is the field strength of the multipole being considered at the stellar rotation pole, $R_{\ast}$ is the stellar
radius, $r$ is a point external to the star, $P_{\ell1}(\cos{\theta})$ and $P_{\ell}(\cos{\theta})$ are the $m=1$ $\ell$th associated Legendre function 
and the $\ell$th Legendre polynomial respectively; $\ell$ represents
the multipole being considered, for example, $\ell=1$ is the dipole and $\ell=2$ the quadrupole. I define the associated Legendre 
functions as 
\begin{equation}
P_{\ell m}(x)=(1-x^2)^{m/2}\frac{d^m}{dx^m}P_\ell(x),
\label{as_le}
\end{equation}
where the Legendre polynomials are
\begin{equation}
P_\ell(x)=\frac{1}{2^\ell\ell!}\frac{d^\ell}{dx^\ell}\left[ (x^2-1)^\ell\right],
\end{equation}
and $x = \cos{\theta}$. The associated Legendre functions are defined without the Condon-Shortley phase (a term of the form $(-1)^m$), as in Ref.~\onlinecite{gre10}.
The basic properties of the associated Legendre functions and the Legendre polynomials can be found in many texts, 
for example Ref.~\onlinecite{kre99}. Their functional form as pertains to the spherical magnetic field 
components is discussed in the appendix.\cite{foot1}

\subsection{The effects of stellar outflows}\label{comps_part2}
Equations (\ref{Br}) and (\ref{Btheta}) do not take into account the influence of outflows on the structure of large scale stellar magnetospheres. At some
distance above the stellar surface, the coronal plasma distorts and pulls open the closed field line loops to produce a stellar wind. To account 
for this transition from the closed coronal field to the open interplanetary magnetic field for the Sun, Refs.~\onlinecite{alt69} and \onlinecite{sch69}
introduced the potential field source surface model. A potential field is one which is current free, that is, the current density $\Jv=\mathbf{0}$ 
everywhere within the stellar magnetosphere, and the field $\Bv$ can be written in terms of the gradient of a magnetostatic scalar potential $\Psi$, that is, 
$\Bv=-\grad\Psi$ (see Sec.~\ref{comps_part3}). The source surface is a spherical surface of radius $R_s$ at which the magnetic 
field is purely radial ($B_{\theta}(R_s)=B_{\phi}(R_s)=0$). This equipotential surface mimics the effect of the coronal plasma, which pulls open the 
large scale closed field lines (although the field may become radial inside this radius). In-situ 
satellite observations of the large scale solar heliospheric field within the region of space around the Sun where the solar wind has blown away the gas and dust that 
pervades the rest of interstellar space, suggest that the value $R_s\sim 2.5\,{\rm R}_{\odot}$ is reasonable for the solar corona.

Several papers have examined the validity of potential field source surface model, and have generalized it by including 
current sheets outside of $R_s$ and by considering a non-spherical source surface. 
The distortion of the field by the coronal plasma at the source surface induces a current and invalidates the assumption that the 
large scale magnetosphere can be described as current free. Reference~\onlinecite{sch78} found that the structure of the field
close to the source surface obtained from magnetohydrodynamic (MHD) models that considered the influence of the coronal plasma on the field line shape 
could be reproduced by adoption of a prolate spheroid source surface (with major axis aligned with the solar rotation axis).
The current-free assumption also often fails at reproducing the field topology of
small scale field regions commonly detected on the Sun. However, for stellar observations, such small scale field regions 
remain well below the resolution achievable from the equivalent solar observations, and the current-free source surface model has been used extensively 
to produce three-dimensional models of the large-scale solar corona and stellar coronae via field extrapolation from observationally derived surface 
magnetic maps.\cite{jar02,gre06,jar08,gre08} Reference~\onlinecite{ril06} found that the large-scale field structure 
of the solar corona is adequately reproduced by the current-free source surface model in the sense that it
produces magnetic field topologies that are similar to those obtained from more complex and computationally intensive MHD 
field extrapolations. Reference~\onlinecite{rom10} also found from their MHD simulations that the global structure of stellar magnetospheres
show little departure from a potential state when evolved over several stellar rotations.
The source surface boundary condition has thus been successfully applied to construct models of the magnetospheres of stars of 
various ages and spectral types,\cite{types,jar02,don06,gre08} as well as planetary magnetospheres.\cite{sch96} Detailed comparisons 
between potential field source surface models and MHD models of global magnetic field topologies can be 
found in Refs.~\onlinecite{ril06} and \onlinecite{gre10}.

\subsection{The magnetic field components with a source surface}\label{comps_part3}
For a spherical source surface of radius $R_s$ the field components, Eqs.~(\ref{Br}) and (\ref{Btheta}), are modified [see Eqs.~(\ref{Brsource}) 
and (\ref{Bthetasource})]. In this section Laplace's equation is solved using separation of variables in spherical polar coordinates in the 
region $R_\ast \le r \le R_s$. From the 
solution I derive expressions for the magnetic field components of a multipole of order $\ell$ subject to the 
following conditions: currents within the region external to the star, $r\ge R_{\ast}$, and interior to the source surface, $r\le R_S$, can be neglected, and 
therefore the field $\Bv$ can be expressed as $\Bv=-\grad\Psi$; the potential is azimuthally symmetric
($\Psi$ does not depend on $\phi$, and hence $B_\phi=0$ everywhere); the radial field component is specified at $r=R_{\ast}$; and the polar field component 
is zero at the source surface $R_s$, that is, $B_\theta(R_s)=0$.

Details of the derivation of the spherical field components with the source surface boundary condition are given in 
Ref.~\onlinecite{gre10}, and only a brief summary is provided here. The large scale stellar magnetosphere is assumed 
to be static. Maxwell's equations for a magnetostatic field are
\begin{align}
\grad \cdot\Bv &= 0 \label{max1} \\
\grad\times\Bv &= \frac{4\pi}{c}\Jv,\label{max2}
\end{align}
where, assuming that currents can be neglected in the region external to the star, Eq.~(\ref{max2}) reduces to $\grad\times\Bv=\mathbf{0}$. 
Hence we obtain Laplace's equation, $\nabla^2 \Psi = 0$, whose general solution in spherical coordinates can be found in most texts on 
electromagnetism (see, for example, Ref.~\onlinecite{duf68}),
\begin{equation}
\Psi = \sum_\ell \sum_m \left[ a_{\ell m}r^\ell + b_{\ell m} r^{-(\ell+1)}\right]P_{\ell m}(\cos{\theta}){\rm e}^{{\rm i} m\phi} \qquad (R_\ast\le r\le R_s).
\label{Psi_extrap}
\end{equation}
The coefficients $a_{\ell m}$ and $b_{\ell m}$ depend on the boundary conditions. Only the axial multipoles (with $m=0$ and 
hence $B_\phi\propto \partial\Psi/\partial\phi \propto m = 0$) are considered in this paper, for which $a_{\ell0}\equiv a_\ell$, $b_{\ell0}\equiv b_\ell$ 
and from Eq.~(\ref{as_le}) $P_{\ell0}(\cos{\theta})=P_\ell(\cos{\theta})$. One boundary condition is to specify the radial field component at the 
stellar surface. For numerical field extrapolation models this condition comes directly from the observationally derived magnetic surface 
maps.\cite{jar02,gre06,gre08} The radial component of the field across the stellar surface for a particular multipole of order $\ell$ is 
given by Eq.~(\ref{Br}) with $r=R_{\ast}$,
\begin{equation}
B_r(R_{\ast})=B^{\ell,{\rm pole}}_{\ast}P_\ell(\cos{\theta}).\label{bound1}
\end{equation}
The other boundary condition is that the field at the source surface, $r=R_s$, becomes purely radial, that is,
\begin{equation}
B_{\theta}(R_s)=0.\label{bound2}
\end{equation}
The field components for a particular axial multipole of order $\ell$ can be found from $\Bv=-\grad\Psi$ with Eq.~(\ref{Psi_extrap}),
\begin{align}
B_r &= -\left[\ell a_\ell r^{\ell-1} - (\ell+1)b_\ell r^{-(\ell+2)}\right] P_{\ell}(\cos{\theta}) \label{br_extrap} \\
B_\theta &= -\left[a_\ell r^{\ell-1} + b_\ell r^{-(\ell+2)}\right] \frac{\rm d}{\rm d\theta}P_{\ell}(\cos{\theta}). \label{bt_extrap}
\end{align}
From Eq.~(\ref{bt_extrap}) the boundary condition (\ref{bound2}) is satisfied if
\begin{equation}
b_\ell = -a_\ell R_s^{2\ell+1},
\label{almblm}
\end{equation}
which holds for any value of $\ell$.\cite{jar08,gre10} If we substitute Eq.~(\ref{almblm}) into Eq.~(\ref{br_extrap}), and apply Eq.~(\ref{bound1}), 
we can derive expression for $a_\ell$ in terms of $R_{\ast}$ and $R_s$. We then substitute this expression and Eq.~(\ref{almblm}) 
into Eqs.~(\ref{br_extrap}) and (\ref{bt_extrap}) and obtain general expressions for $B_r$ and $B_{\theta}$ for the large scale 
magnetosphere with a source surface,
\begin{align}
B_r &= B^{\ell,{\rm pole}}_{\ast} \left(\frac{R_{\ast}}{r}\right)^{\ell+2} P_\ell(\cos{\theta}) \left[\frac{\ell r^{2\ell+1}+(\ell+1)R_s^{2\ell+1}}{\ell R_{\ast}^{2\ell+1}+(\ell+1)R_s^{2\ell+1}} \right] \label{Brsource}, \\
B_\theta &= \frac{B^{\ell,{\rm pole}}_{\ast}}{\ell+1}\left(\frac{R_{\ast}}{r}\right)^{\ell+2}P_{\ell1}(\cos{\theta}) \left[\frac{-(\ell+1)r^{2\ell+1}+(\ell+1)R_s^{2\ell+1}}{\ell R_{\ast}^{2\ell+1}+(\ell+1)R_s^{2\ell+1}}\right]\label{Bthetasource}.
\end{align}
for $R_\ast\le r\le R_s$. To derive Eq.~(\ref{Bthetasource}) we used the fact that $P_{\ell1}(\cos{\theta})=-\rm d P_\ell(\cos{\theta})/\rm d\theta$ 
(see Eq.~(\ref{as_le}) with $m=1$ and $x=\cos{\theta}$). The field components describe the structure of the magnetic field in the region exterior 
to the star/planet ($r\ge R_{\ast}$) and interior to the source surface ($r\le R_s$). At radii beyond $R_s$, as for the Sun, the field is usually 
described by a Parker spiral\cite{par58} (the large scale solar field being dragged into a spiral shape by the outflowing wind), which is not 
considered in this paper. As $R_s$ is decreased, more of the closed field line loops are converted to open field lines, and in the limit 
$R_s\to\infty$, Eqs.~(\ref{Brsource}) and (\ref{Bthetasource}) reduce to Eqs.~(\ref{Br}) and (\ref{Btheta}). From the modified field components 
it is possible to derive the equation of the field lines.

\section{Equation of the field lines}\label{equ}
The differential equation for the field lines of an arbitrary magnetic multipole $\ell$ is\cite{wil87,wil88,gre06}
\begin{equation}
\frac{B_r}{dr}=\frac{B_{\theta}}{rd\theta}=\frac{B_{\phi}}{r\sin{\theta}d\phi}=\frac{B}{ds},
\label{path}
\end{equation}
where $s$ is a coordinate along the path of the particular field line under consideration, and $B=|\Bv|=(B_r^2+B_{\theta}^2+B_{\phi}^2)^{1/2}$ 
is the magnitude of the field. For axial multipoles the equation describing the path of the field lines can be determined by solving the first two 
terms of Eq.~(\ref{path}).\cite{gre10} Likewise, the equation for the field lines of a point electric linear multipole can be found in an analogous 
manner, see for example Appendix~A of Ref.~\onlinecite{gra09}, either in spherical or Cartesian coordinates.\cite{smy50,kir85}

We substitute Eqs.~(\ref{Brsource}) and (\ref{Bthetasource}) into Eq.~(\ref{path}) and rearrange terms to obtain
\begin{equation}
\frac{(\ell+1)P_\ell(\cos{\theta})d\theta}{P_{\ell1}(\cos{\theta})} = \frac{dr}{r}
\left[ \frac{-(\ell+1)r^{2\ell+1}+(\ell+1)R_ s^{2\ell+1}}{\ell r^{2\ell+1}+(\ell+1)R_s^{2\ell+1}}\right].
\label{fieldline_equ}
\end{equation}
To obtain the equation of a particular field line that has a maximum radial extent of $r_{\max}$ (see Fig.~\ref{fig0}) Eq.~(\ref{fieldline_equ}) 
must be integrated from $r_{\max}$ to some point $r$ along the loop where $R_\ast\le r\le r_{\max}$ and $R_\ast < r_{\max} \le R_s$. 
For closed field lines at $r=r_{\max}$, $B_r=0$, and the field only has a $B_{\theta}$ component (see Fig.~\ref{fig0}). For open field lines 
$r_{\max}=R_s$. The right-hand side of Eq.~(\ref{fieldline_equ}) is integrated by writing the integral as
\begin{equation}
\left[\frac{-(\ell+1)}{\ell(2\ell+1)}\ln{(\ell r^{2\ell+1}+\beta)}\right]^r_{r_{\max}} + \!\int_{r_{\max}}^r \frac{\beta r^{2\ell}dr}{r^{2\ell+1}(\ell r^{2\ell+1}+\beta)}, \label{rhs1}
\end{equation}
where $\beta = (\ell+1)R_s^{2\ell+1}$. The inclusion of the additional $r^{2\ell}$ term in the second integral allows it to be simplified by the 
substitution $\alpha =\ell r^{2\ell+1}+\beta$. The second term can be further simplified by partial
fraction decomposition and directly integrated to give
\begin{equation}
\frac{1}{\ell}\ln{\left[ \left(\frac{r}{r_{\max}}\right)^\ell \left(\frac{\ell r_{\max}^{2\ell+1}+(\ell+1)R_s^{2\ell+1}}{\ell r^{2\ell+1}+(\ell+1)R_s^{2\ell+1}} \right) \right]}.
\label{rhs_final}
\end{equation}

The left-hand side of Eq.~(\ref{fieldline_equ}) is more complicated to integrate. We want to integrate from the maximum radial extent of a given field 
line to some point along the field line at $r$. I assume that at $r_{\max}$, $\theta = \Theta$, and the left-hand side of Eq.~(\ref{fieldline_equ}) is
\begin{equation}
\int_{\Theta}^{\theta} \frac{(\ell+1)P_\ell(\cos{\theta})}{P_{\ell1}(\cos{\theta})}d\theta.
\label{lhs}
\end{equation}
Equation (\ref{lhs}) can be integrated following the arguments in Ref.~\onlinecite{wil87}. However, given the differing
definitions of the associated Legendre functions, and the fact that I also wish to evaluate the integration constant, it is instructive to
briefly summarize the method here. From Ref.~\onlinecite{erd53}, Eq.~(15), the Legendre polynomials are related to the 
$P_{\ell1}(\cos{\theta})$ and $P_{(\ell-1)1}(\cos{\theta})$ associated Legendre functions as
\begin{equation}
P_\ell(\cos{\theta})=\frac{\cos{\theta}}{\ell\sin{\theta}}P_{\ell1}(\cos{\theta}) - \frac{1}{\ell\sin{\theta}}P_{(\ell-1)1}(\cos{\theta}).
\end{equation}
Using this relation Eq.~(\ref{lhs}) becomes
\begin{equation}
\int^{\theta}_{\Theta}\left[ \left(\frac{\ell+1}{\ell} \right)\frac{\cos{\theta}}{\sin{\theta}}-\left(\frac{\ell+1}{\ell}\right)\frac{1}{\sin{\theta}}\frac{P_{(\ell-1)1}(\cos{\theta})}{P_{\ell1}(\cos{\theta})}\right]d\theta.
\label{lhs2}
\end{equation}
The first term can be integrated directly, but the second term cannot. From Ref.~\onlinecite{erd53}, Eq.~(19),
\begin{equation}
-\sin{\theta}\frac{dP_{\ell1}(\cos{\theta})}{d\theta} = -\ell\cos{\theta}P_{\ell1}(\cos{\theta}) + (\ell+1)P_{(\ell-1)1}(\cos{\theta}).
\label{relation1}
\end{equation}
I also note that
\begin{equation}
\frac{1}{\sin{\theta}}\frac{d}{d\theta}\left[ \sin^2{\theta}P_{\ell1}(\cos{\theta})\right]=2\cos{\theta}P_{\ell1}(\cos{\theta})+\sin{\theta}\frac{dP_{\ell1}(\cos{\theta})}{d\theta}.
\label{relation2}
\end{equation}
By combining Eqs.~(\ref{relation1}) and (\ref{relation2}) an expression for $(\ell+1)P_{(\ell-1)1}(\cos{\theta})$ can be obtained, which, 
upon substitution into Eq.~(\ref{lhs2}), and some manipulations, yields,
\begin{equation}
\int_{\Theta}^{\theta}\left[-\frac{1}{\ell}\frac{\cos{\theta}}{\sin{\theta}}+ \frac{1}{\ell}\frac{\dfrac{d}{d\theta}[ \sin^2{\theta}P_{\ell1}(\cos{\theta})]}{\sin^2{\theta}P_{\ell1}(\cos{\theta})} \right] d\theta.
\label{express}
\end{equation}
Equation~\eqref{express} can now be integrated directly,
\begin{equation}
\frac{1}{\ell}\ln{\left| \frac{\sin{\theta}P_{\ell1}(\cos{\theta})}{\sin{\Theta}P_{\ell1}(\cos{\Theta})}\right|}.
\label{lhs_final}
\end{equation}

\begin{figure}[h!]
\centering
\scalebox{0.45}{\includegraphics{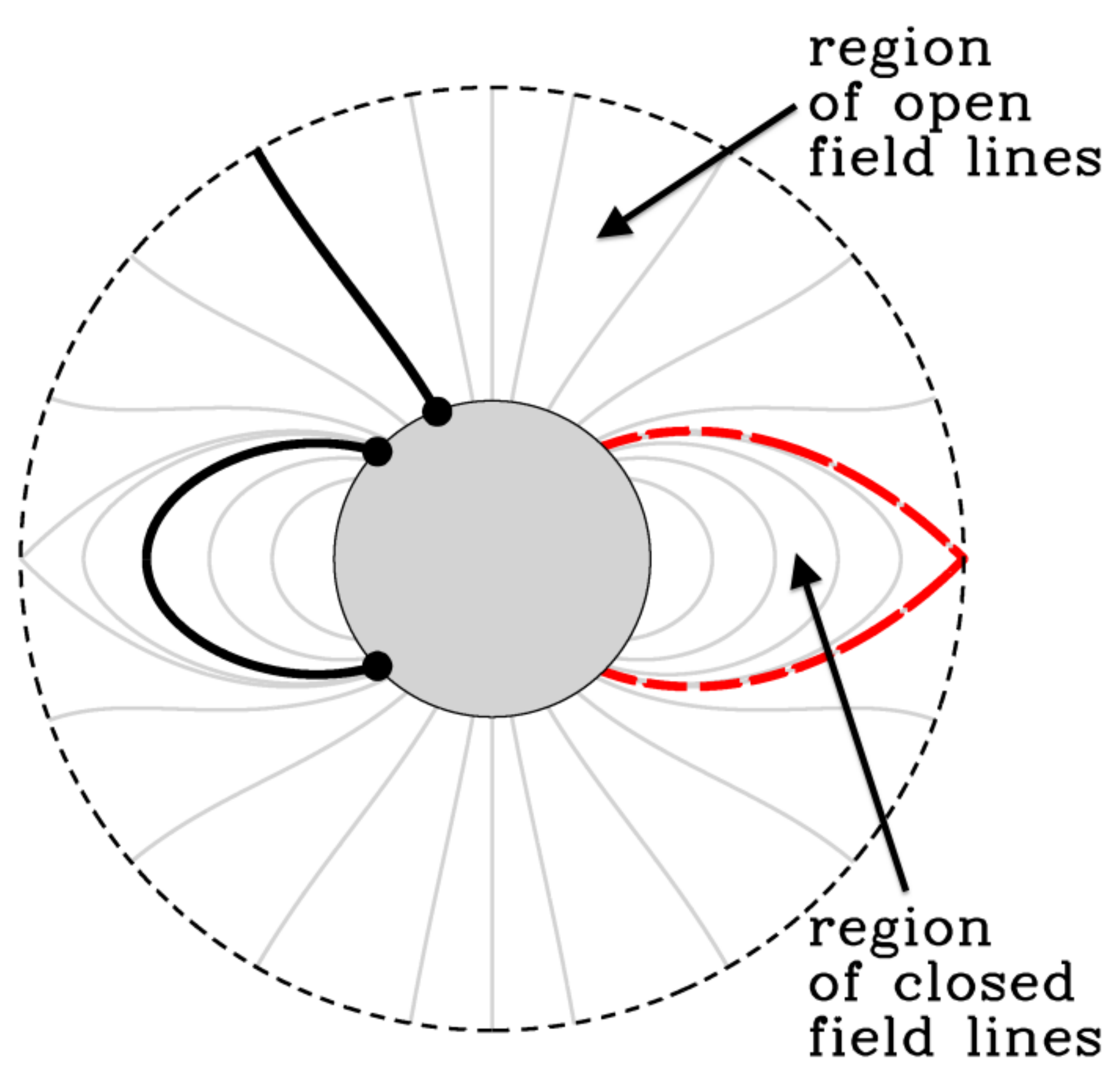}}
\caption{\label{fig_colored_dipole}(color online) A star of radius $R_\ast$, shaded gray, with an axial dipole magnetic field with a source surface 
(the large dashed circle) of radius $R_s$. The source surface allows for the inclusion of regions of open field lines, along which a stellar wind 
is launched, in addition to the regions of closed field lines. Each closed field region is surrounded by regions of open field lines. The field lines 
are shown in light gray. The solid black lines illustrate particular field lines, with their footpoints on the stellar surface denoted by the solid black 
circles. Open field lines have a single footpoint, and closed field lines have two footpoints on the star. The field lines within each region of closed 
field lines reach their maximum radial extent of $r_{\max}$ along a line of constant polar angle. Different values of $r_{\max}$ correspond to 
different field lines within each closed field region. The long-dashed line denotes the closed field line loop with $r_{\max}=R_s$. The footpoints 
of this closed field line represent the boundary at the stellar surface between regions of closed and open field.}
\end{figure}

\begin{figure}[h!]
\centering
\scalebox{0.25}{\includegraphics{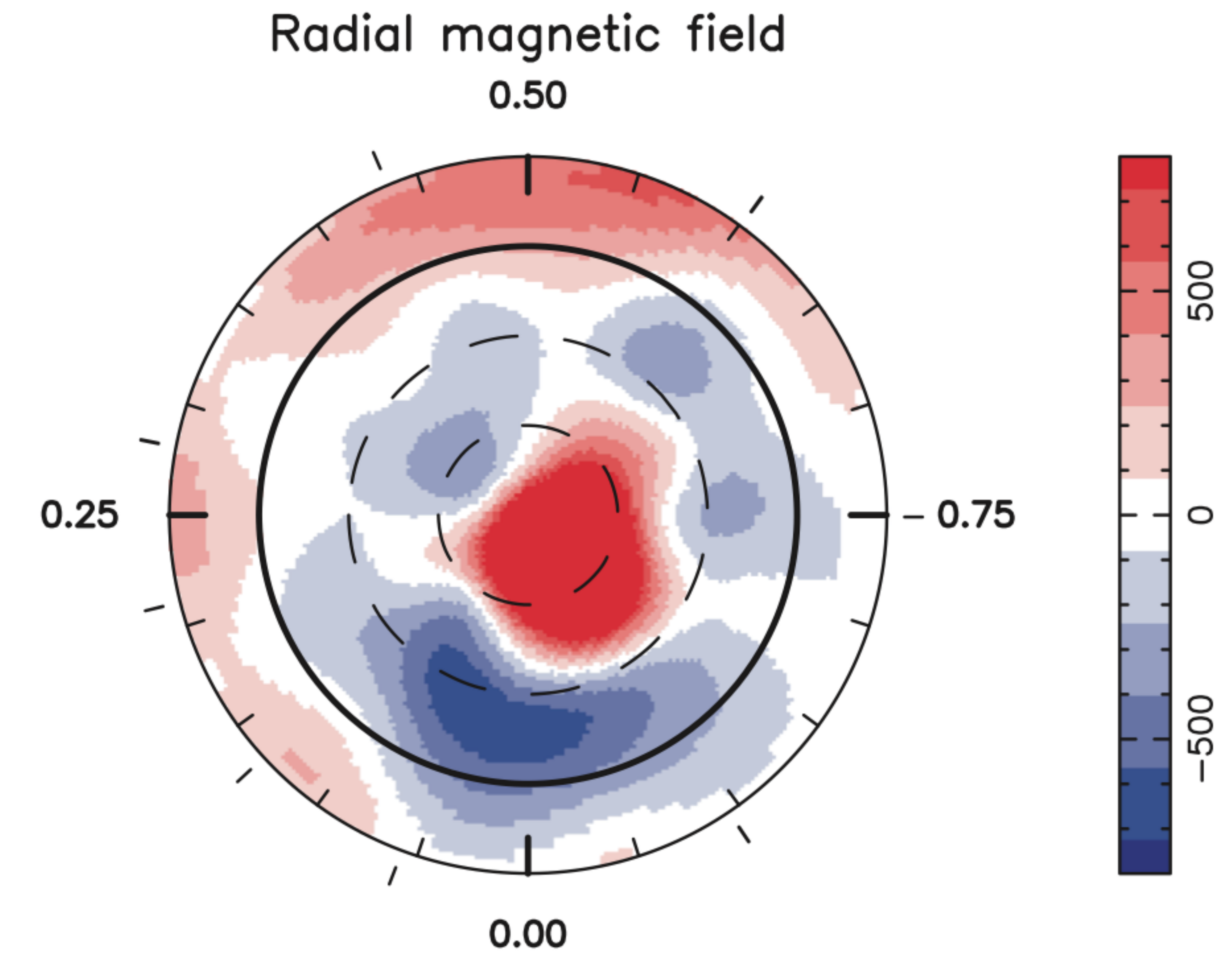}}
\scalebox{0.35}{\includegraphics{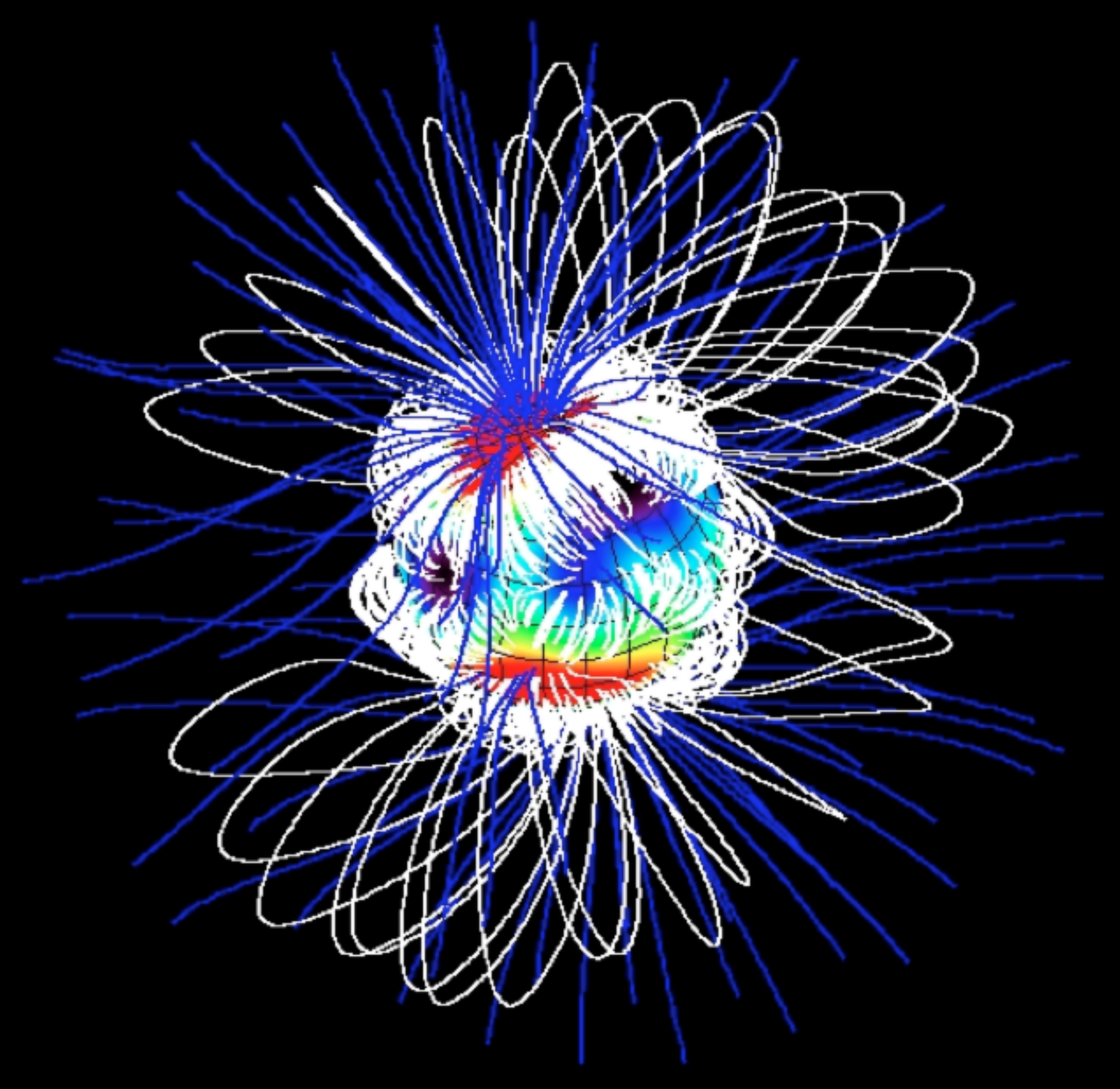}}
\caption{\label{fig_map}(color online) (a) A flattened polar projection showing the radial field component of the surface magnetic field of the 
forming star V2129\,Oph ($\sim1.35$\,M$_{\odot}$).\cite{don07} The stellar equator is shown as the bold circle, with lines of constant latitude 
separated by 30$^{\circ}$ as the dashed lines.  There is a positive field spot slightly offset from the pole, surrounded by a ring of negative field,
which itself is surrounded by a ring of positive field below the equator.  Fluxes 
given in Gauss. Numbers and tick marks around the circumference denote the rotation phase and phases of observation respectively. The large 
scale magnetic field is found to be dominantly octupolar, which is apparent from the (b) extrapolation of the three-dimensional coronal magnetic 
field, which is constructed from the magnetic map. Closed field lines are shown in white, with open field lines, along which a stellar wind could be 
launched, shown in blue (see the online version for colors). The large scale field approximately resembles a tilted dipole, and the medium scale 
field resembles an octupole with three rings of closed field. Details of the observational techniques used to construct stellar magnetic maps and 
the numerical field extrapolation model can be found in Refs.~\onlinecite{don09} and \onlinecite{gre10}, respectively.}
\end{figure}

\begin{figure}[h!]
\centering
\scalebox{0.3}{\includegraphics{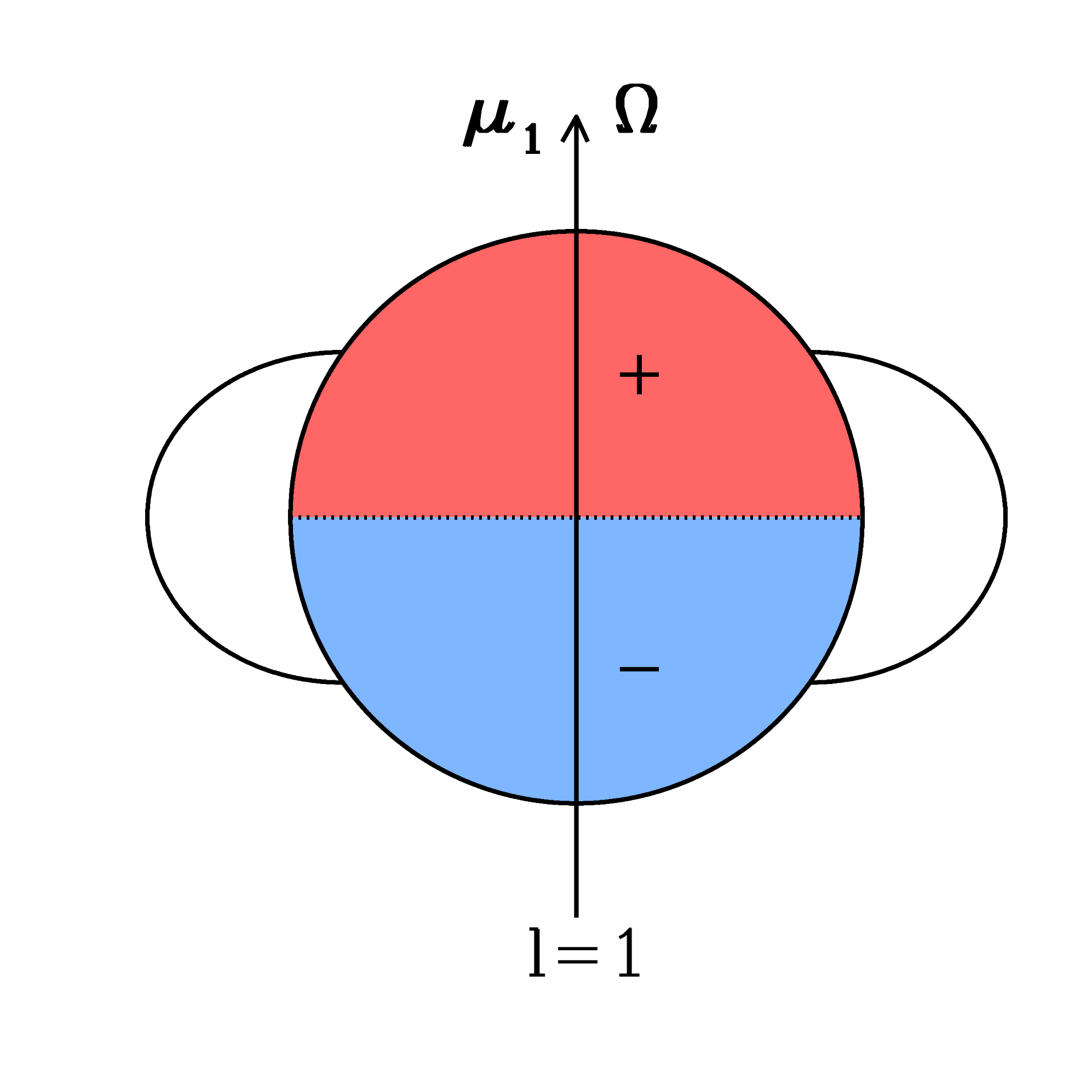}}
\scalebox{0.3}{\includegraphics{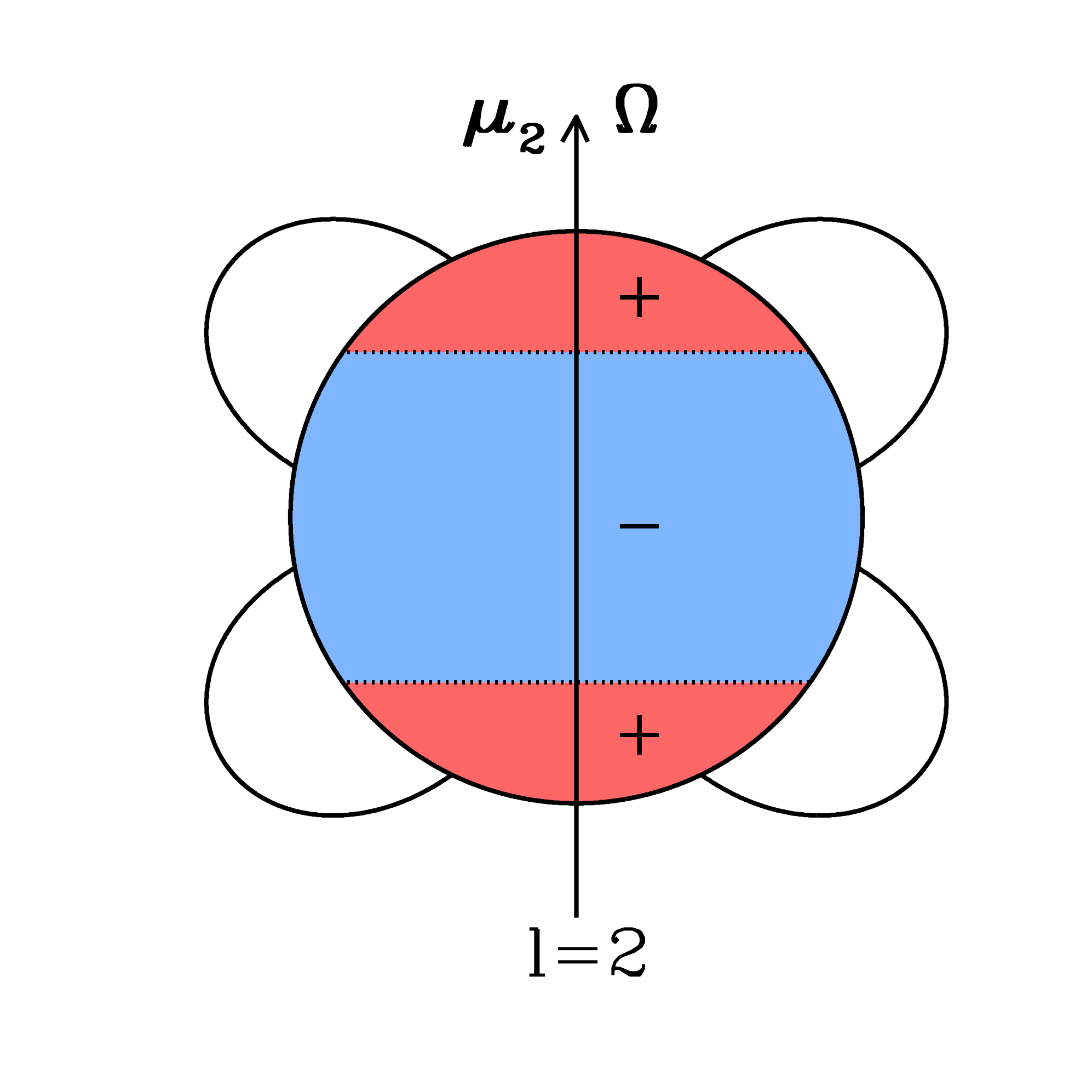}}
\scalebox{0.3}{\includegraphics{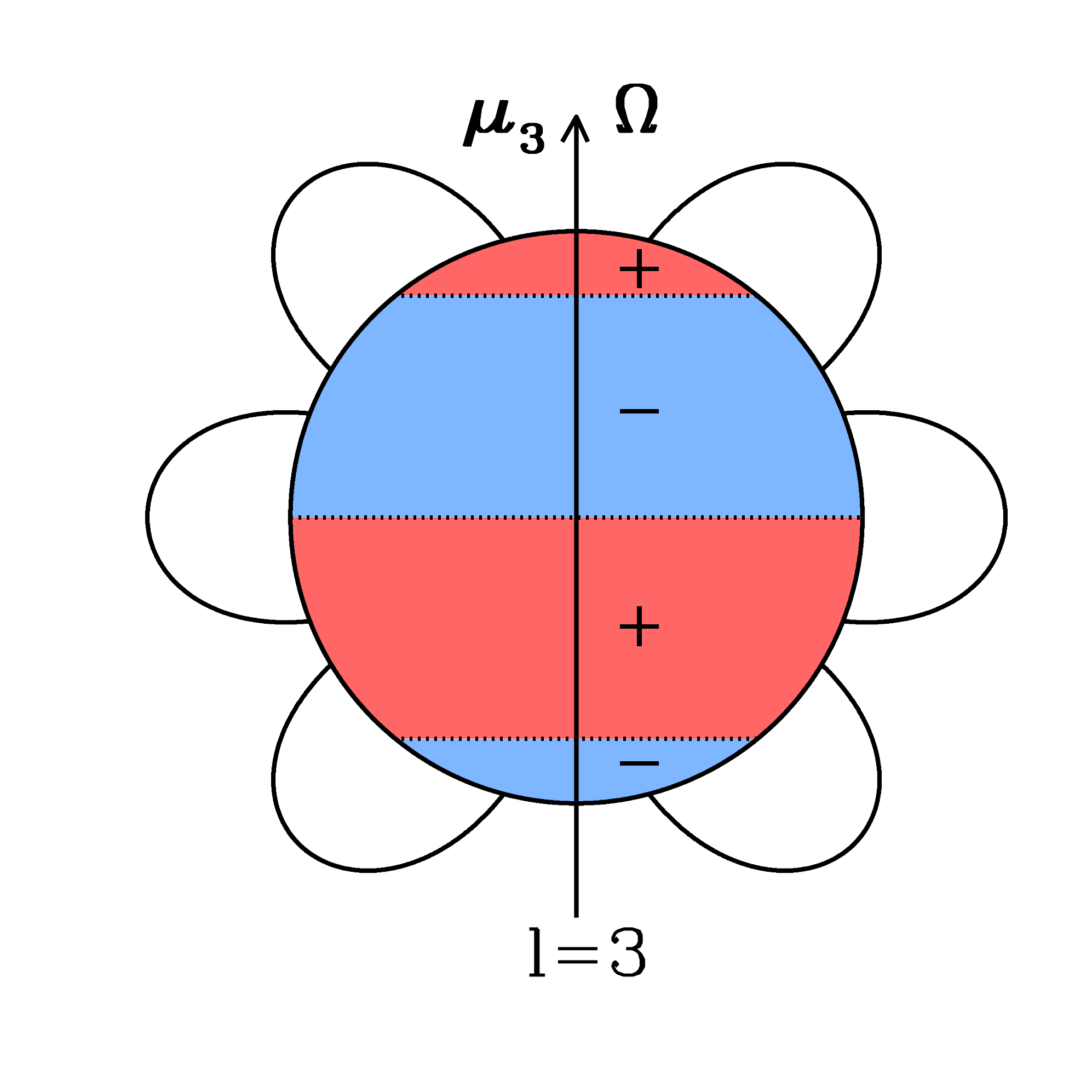}}
\caption{\label{color_plots}(color online) The first three lowest order multipoles, the dipole ($\ell=1$), the quadrupole ($\ell=2$) and the octupole 
($\ell=3$). The multipole moment symmetry axis, denoted by $\boldsymbol{\upmu}_1$ for the dipole, $\boldsymbol{\upmu}_2$ for the 
quadrupole, and $\boldsymbol{\upmu}_3$ for octupole, is assumed to be aligned with the stellar rotation axis, denoted by $\mathbf{\Omega}$. 
The plus/minus signs denote regions of positive/negative field. The order of a multipole $\ell$ is the number of polarity changes in the surface 
field between the north and south pole of the star along meridians (lines of constant longitude); $\ell$ is the number of roots of 
$P_\ell(\cos{\theta})$ between $\theta=0$ and $\theta=\pi$. A multipole of order $\ell$ has $2\ell$ regions of closed loops around the 
entire star in each meridional plane (planes with $\phi={\rm constant}$). In three dimensions the closed field regions form rings of closed 
field around the star, with $\ell$ rings for a multipole of order $\ell$.}
\end{figure}

\begin{figure}[h!]
\centering
\scalebox{0.80}{\includegraphics{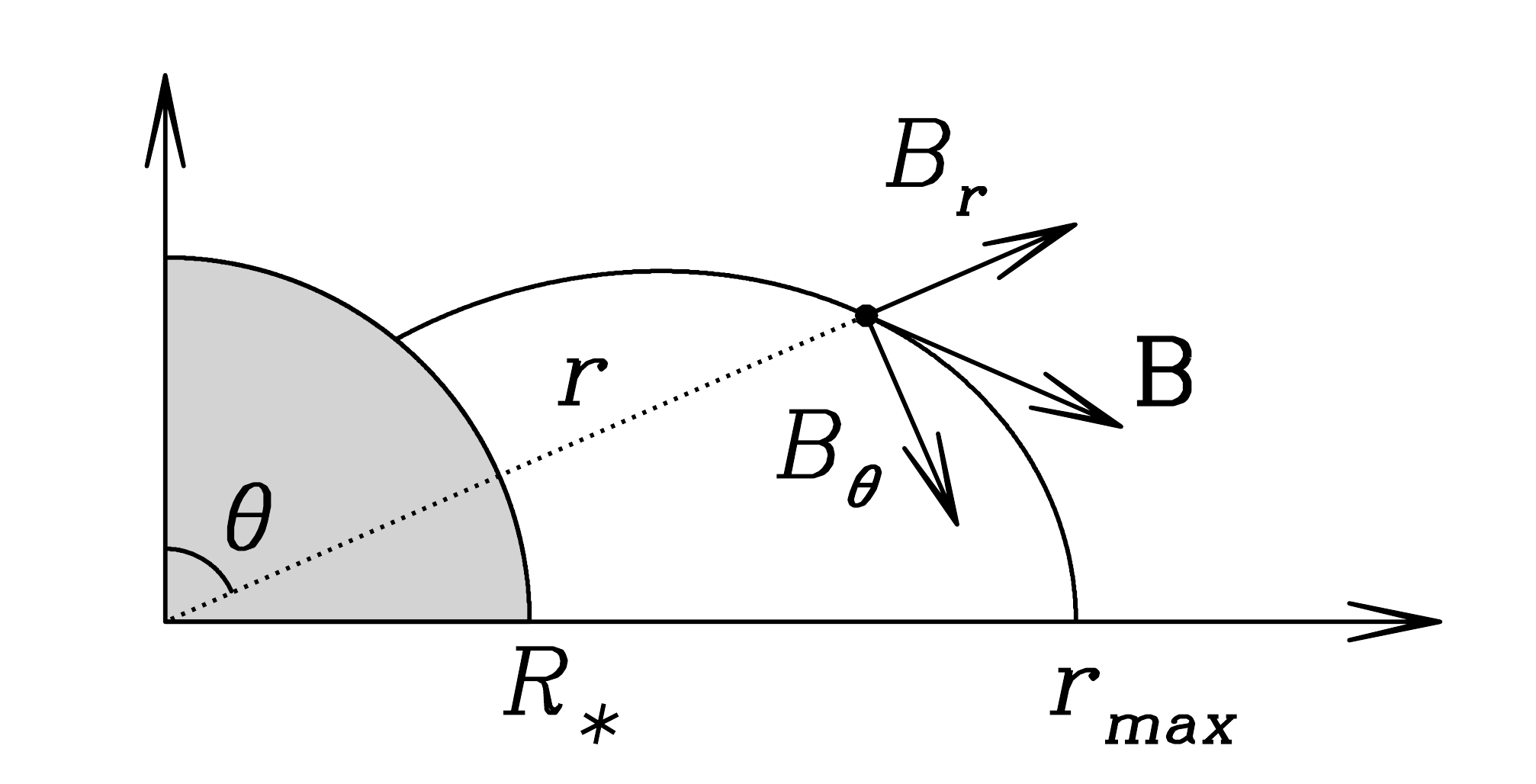}}
\caption{\label{fig0}A field vector $\Bv$ decomposed into the radial $B_r$ and polar $B_\theta$ components at a point along a field line a distance $r$ 
from the center of the star (shaded gray) at a co-latitude of $\theta$. The field components are used to illustrate their definitions 
and are not to scale. The field line reaches a maximum radial extent of $r_{\max}$. Only the first quadrant is shown because the magnetic 
fields considered in this paper are reflectionally symmetric in the horizontal axis and rotationally symmetric about the vertical axis.}
\end{figure}

The modulus signs can be removed as the signs of $P_{\ell1}(\cos{\theta})$, $P_{\ell1}(\cos{\Theta})$, $\sin{\theta}$, and $\sin{\Theta}$ are always 
the same (see the appendix). Finally, by combining Eqs.~(\ref{rhs_final}) and (\ref{lhs_final}), the solution of Eq.~(\ref{fieldline_equ}) which is the 
equation of the field lines of an axial multipole with a source surface is found to be
\begin{equation}
\frac{\bar{r}^\ell}{l\bar{r}^{2\ell+1}+(\ell+1)\bar{R}_s^{2\ell+1}}=\frac{\bar{r}_{\max}^\ell}{(\ell\bar{r}_{\max}^{2\ell+1}+(\ell+1)\bar{R}_s^{2\ell+1})} \frac{\sin{\theta}P_{\ell1}(\cos{\theta})}{\sin{\Theta}P_{\ell1}(\cos{\Theta})},
\label{final}
\end{equation}
where $\bar{r}=r/R_{\ast}$, $\bar{R}_s=R_s/R_{\ast}$ and $\bar{r}_{\max}=r_{\max}/R_{\ast}$. Although Eq.~\eqref{final} may lack simplicity, once a 
source surface location and the maximum radius of the field line of interest has been chosen, the right-hand side of Eq.~(\ref{final}) simplifies to 
${\rm constant} \times\sin{\theta}P_{\ell1}(\cos{\theta})$. Different field lines correspond to different values of this constant. Each region of closed 
field lines (see Fig.~\ref{fig_colored_dipole}) must be considered separately because each region of closed loops reach their maximum radial extent 
at a different value of $\Theta$. For closed field line loops $\Theta$ is the value
of $\theta$ where $P_\ell(\cos{\theta})=0$ (and consequently $B_r=0$ (see the appendix). For open field lines $r_{\max}=R_s$, and $\Theta$ is the 
polar coordinate where a particular open field line crosses the source surface.

The footpoints of a particular closed field line loop, that is, the co-latitudes $\theta_{\ast}$ where the closed loop attaches to the stellar 
surface (see Fig.~\ref{fig_colored_dipole}), are obtained by integrating Eq.~(\ref{fieldline_equ}) from $r_{\max}$ to $R_{\ast}$,
which is equivalent to replacing $\bar{r}$ by $1$ in Eq.~(\ref{final}),
\begin{equation}
\sin{\theta_{\ast}}P_{\ell1}(\cos{\theta_{\ast}})=\left(\frac{1}{\bar{r}_{\max}}\right)^\ell \left(\frac{\ell\bar{r}_{\max}^{2\ell+1}+(\ell+1)\bar{R}_s^{2\ell+1}}{\ell+(\ell+1)\bar{R}_s^{2\ell+1}}\right)\sin{\Theta}P_{\ell1}(\cos{\Theta}).
\label{final_star}
\end{equation}
There are two solutions $\theta_\ast$ of Eq.~\eqref{final_star} corresponding to the two footpoints of the particular closed field line loop under consideration.

The location of the source surface $\bar{R_s}$ controls the amount of open field relative to the closed field. Of particular interest is the amount of open 
flux through the stellar surface. By integrating Eq.~(\ref{fieldline_equ}) from $r_{\max}=R_s$ to $r=R_{\ast}$ [achieved by substitution of 
$\bar{r}_{\max}=\bar{R}_s$ and $\bar{r}=1$ into Eq.~(\ref{final})] it is possible to determine the location of the boundaries between regions of open and
closed field lines at the stellar surface which is where the largest closed field line loop connects to the star (for example, see the long-dashed field line
in Fig.~\ref{fig_colored_dipole}),
\begin{equation}
\sin{\theta_{\ast}}P_{\ell1}(\cos{\theta_{\ast}})=\frac{(2\ell+1)\bar{R}_s^{\ell+1}}{\ell+(\ell+1)\bar{R}_s^{2\ell+1}}\sin{\Theta}P_{\ell1}(\cos{\Theta}).
\label{final_source}
\end{equation}
From Eq.~\eqref{final_source} it is possible to derive an analytic expression for the open flux through the stellar surface, which is an important 
quantity for models of stellar rotational evolution. Due to angular momentum loss via outflows, the rotation rates of low mass stars slow during 
the hydrogen burning phase of their evolution.\cite{tas97} The angular momentum that can be removed by a stellar wind depends on the amount 
of open flux. 

\subsection{Field line plots}
\begin{figure}[h!]
\centering
\scalebox{0.40}{\includegraphics{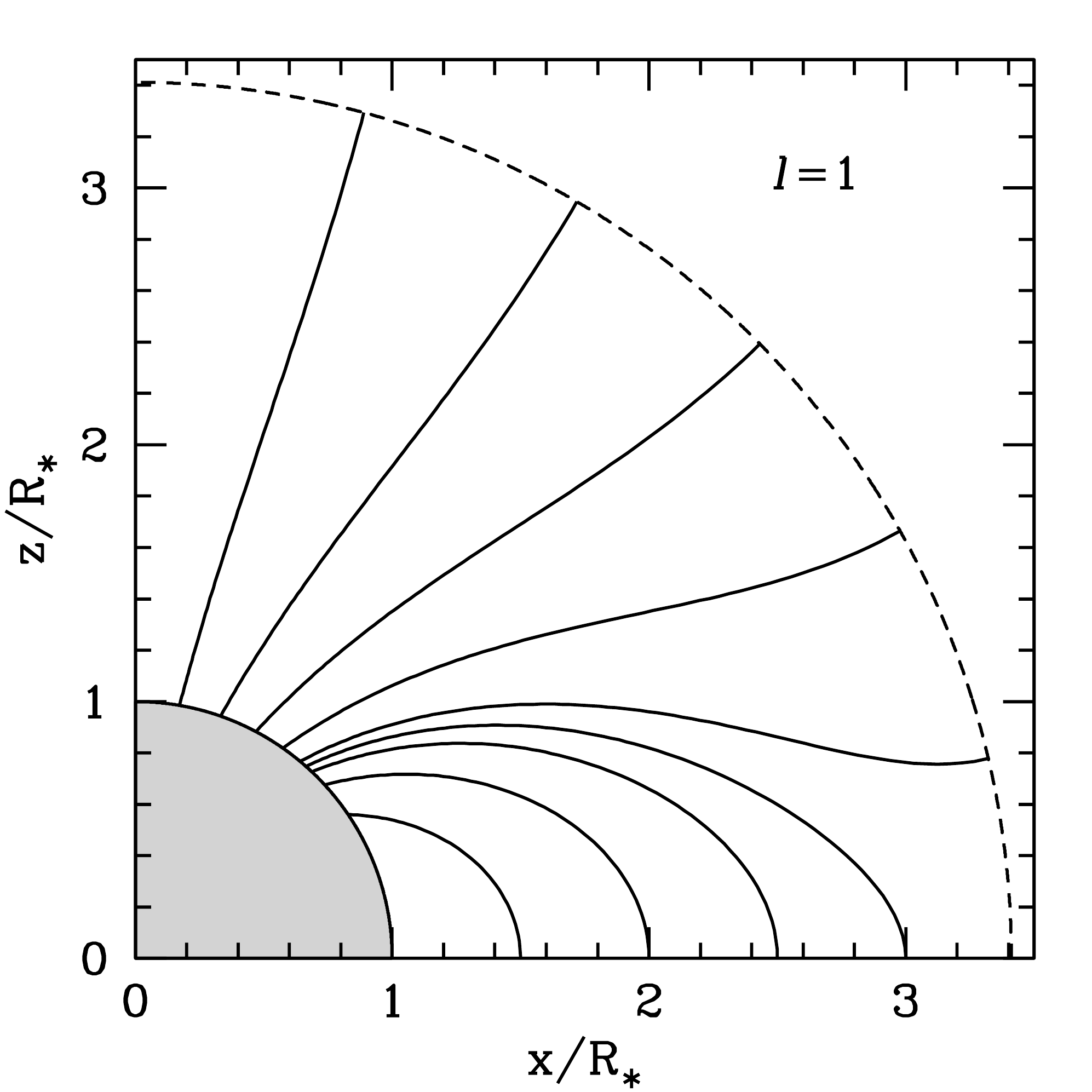}}
\scalebox{0.40}{\includegraphics{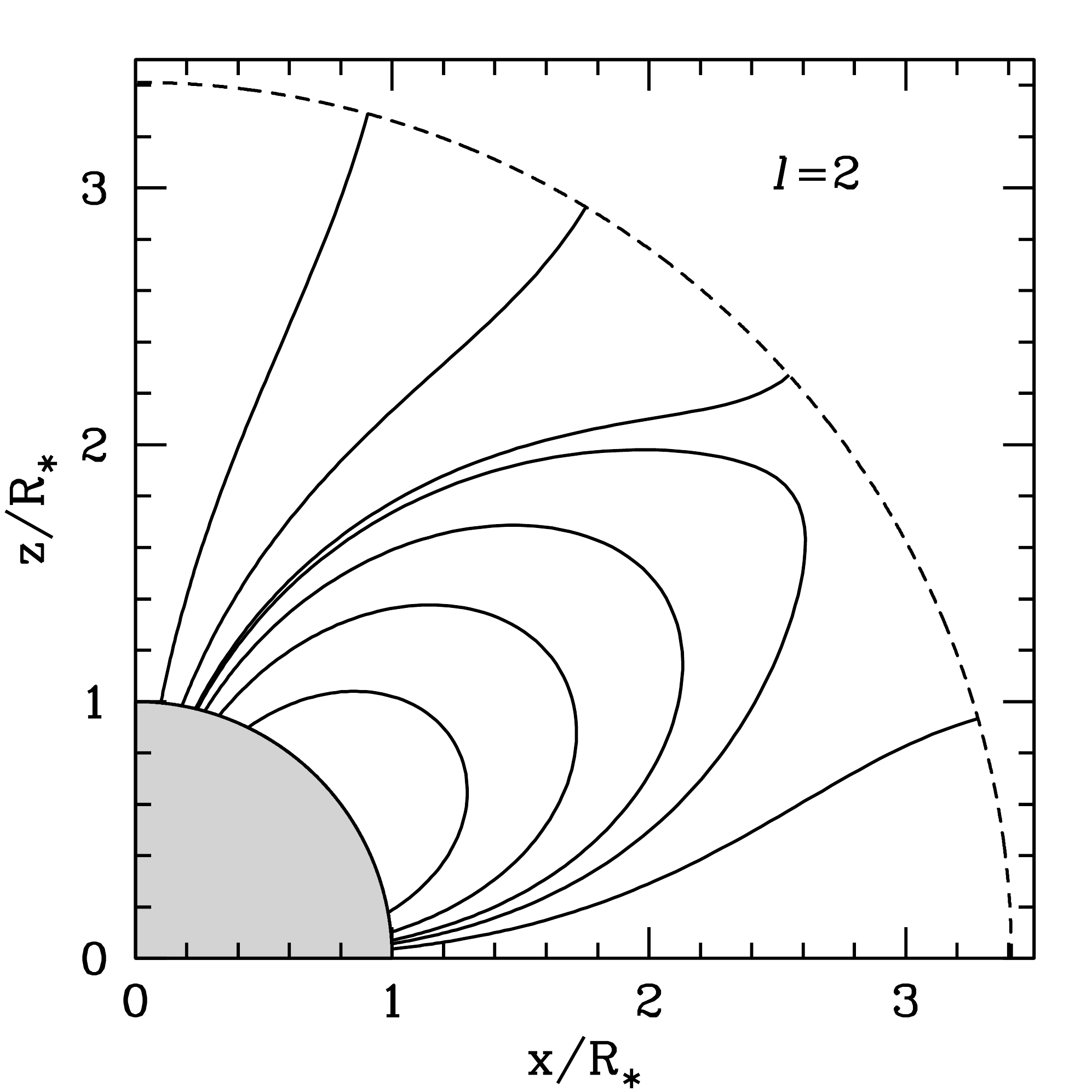}}
\scalebox{0.40}{\includegraphics{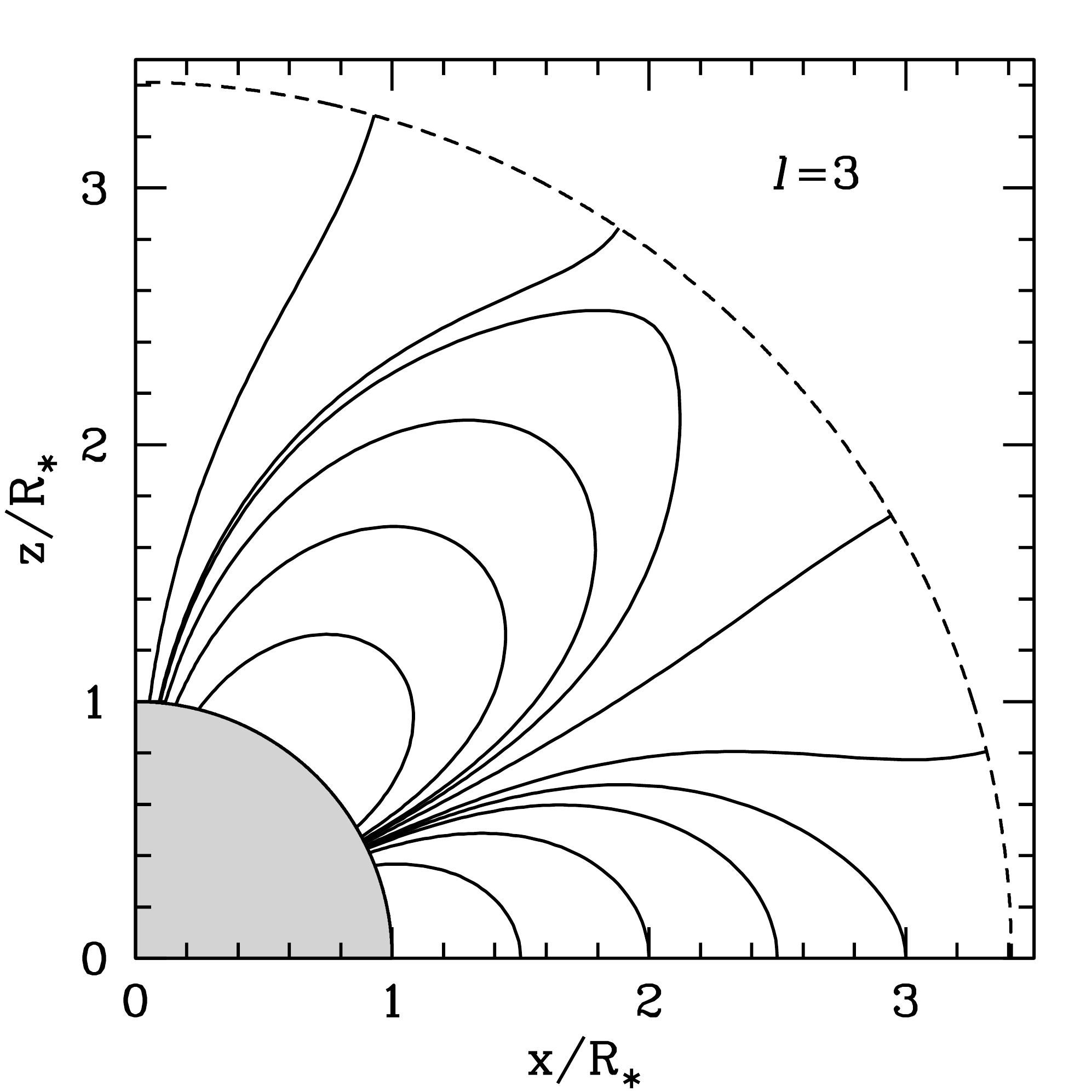}}
\caption{\label{fig1x}(a) The field lines of a dipole ($\ell =1$), (b) a quadrupole ($\ell =2$), and (c) an octupole ($\ell=3$) with a source 
surface at $R_s\sim3.4\,{\rm R}_\ast$ plotted as the dashed line. The star is shaded in gray. The shapes of the closed 
field lines are calculated for each multipole from Eq.~(\ref{final}). The shapes of the 
open field lines are calculated in a similar way from $R_\ast$ to $R_s$ by considering the range of co-latitudes $\theta$ not occupied
by the footpoints of the closed loops. For the quadrupole and the octupole (and other higher order multipoles) there are regions of 
open field at lower latitudes along which a stellar wind could be launched. The magnetic fields are rotationally symmetric about the $z$-axis and 
reflectionally symmetric in the $x$-$y$ plane.}
\end{figure}

The equation of the field lines, Eq.~(\ref{final}), makes possible a visualization of the magnetic field topologies of arbitrary magnetic multipoles 
of order $\ell$, which incorporate regions of both closed and open field lines. Figure~\ref{fig1x} shows the field lines of the three 
lowest order axial multipoles, the dipole ($\ell=1$), the quadrupole ($\ell=2$) and the octupole ($\ell=3$). For a multipole
of order $\ell$ there are $2\ell$ regions of closed field line loops around the entire circumference of the star in each meridional plane 
(planes with $\phi={\rm constant}$). Additional properties of axial multipoles are discussed in the appendix. The path of the field lines 
illustrated in Fig.~\ref{fig1x} are obtained by solving Eq.~(\ref{final}) as follows. Suppose we are interested in a particular closed
field line loop that has footpoints $\theta_1$ and $\theta_2$, and which reaches its maximum radial extent at $\Theta$. The footpoints 
$\theta_1$ and $\theta_2$ are obtained by solving Eq.~(\ref{final_star}). For a given region of closed field lines for an axial multipole of order 
$\ell$ and given values of $\bar{R}_s$ and $\bar{r}_{\max}$, the right-hand side of Eq.~(\ref{final_star}) is a constant, say $\eta$. The 
footpoints of the closed field line $\theta_{\ast}$ can then be found by finding the roots of the function 
$f(\theta_{\ast})=\sin{\theta_{\ast}P_{l,1}(\cos{\theta}_{\ast})}-\eta$ using a standard root finding algorithm. There are two roots 
$\theta_{\ast}$ corresponding to the footpoints $\theta_1$ and $\theta_2$. Because axial multipoles are rotationally symmetric about 
the $z$-axis and reflectionally symmetric in the $x$-$y$ plane, I only consider values of $\theta$ between zero and $\pi/2$. 
Odd $\ell$ multipoles have a region of closed field lines that span the stellar equatorial plane,
$\Theta=\pi/2$ (see Fig.~\ref{fig1x}), and for this region there is only one solution $\theta_{\ast}$ of Eq.~(\ref{final_star}) 
for $0\le\theta\le\pi/2$. Once the $\theta_1$ and $\theta_2$ of a closed loop have been determined, it is straightforward to solve for the 
path of the field line by solving Eq.~(\ref{final}), again using a root finding technique to find $\bar{r}$ at each value of $\theta$ 
between $\theta_1$ and $\theta_2$. Equation~(\ref{final}) can also be used to trace the paths of open field lines. For open field lines, 
$\bar{r}_{\max}$ is replaced by $\bar{R}_s$ in Eq.~(\ref{final}), and $\Theta$ is the polar angle where a particular open field line arrives 
at the source surface. For example, for a particular multipole that has closed field lines which reach their maximum radial extent at 
$\Theta_i$, where $i=1,2,\ldots$, the open field lines arrive at the source surface ($r=R_s$)
at all angles between $0$ and $\pi/2$ except $\Theta_i$.

\section{Summary}\label{sum}
Over the past five years the current generation of optical spectropolarimeters has allowed stellar magnetic field topologies to be probed in unprecedented
detail. Large observational programs have provided maps of the magnetic
fields of stars spanning a range of masses, rotation rates, and evolutionary stages.\cite{don09} In particular,
magnetic maps of forming solar-like stars can now be obtained.\cite{don07} The large scale field topologies vary from simple dipoles to more complex magnetic
fields consisting of several multipole components.\cite{gre10} Given the availability of the observational
datasets, new models that incorporate magnetic fields are under development such as models that consider how the magnetospheres of forming solar-like stars interact
with their surrounding planet-forming disks.\cite{gre08,gre10,rom10}

I have discussed how the solution
of Laplace's equation using separation of variables can be applied to construct models of stellar magnetospheres where the influence
of outflows on the global field structure is considered. With the assumption that currents have a negligible effect on the large scale structure of 
stellar magnetic fields in the region between the stellar surface and the source surface, the field components have been derived by solving the 
Laplace boundary value problem, thus by passing the more complicated problem of solving Poisson's equation. Basic properties of multipole 
magnetic fields have been discussed, and from the field components, Eqs.~(\ref{Brsource}) and (\ref{Bthetasource}), an analytic equation 
for the field lines of an arbitrary axial multipole with a source surface has been derived. The resulting expression, Eq.~(\ref{final}), is straightforward 
to solve using a root finding algorithm and its use only requires knowledge of the roots of the $m=1$ $\ell$th associated Legendre function 
$P_{\ell1}(\cos{\theta})$.

I have concentrated on deriving the equation of the field lines for a modified magnetic multipole in spherical coordinates
by solving Eq.~(\ref{path}). Spherical coordinates are the most natural coordinate system to use
in the description of stars and planets and their immediate environments. Other authors have derived analytic expressions
for the field lines of axisymmetric multipoles without the source surface boundary condition in spherical polar coordinates.\cite{wil87,jef88,bac88}
The external field line equation for the magnetic multipoles is identical to that obtained for the field lines of point electric linear multipoles. 
(However, the internal and contact field lines of magnetic and electric multipoles of the same order differ.\cite{gra10})
For example, the path of the electrostatic field lines for a linear quadrupole in spherical coordinates, can be found in Appendix~A of Ref.~\onlinecite{gra09}.
For the electrostatic case it is much more common for the path of the field lines to be determined in Cartesian coordinates. 
Reference~\onlinecite{smy50} provides a Cartesian expression for the field lines of an arbitrary linear electric multipole. Several other overviews 
describe algorithms for simple computer programs that can be adapted to visualize magnetostatic or electrostatic fields.\cite{kri85,kir85,kir86}

The magnetospheric structures considered in this paper have certain limitations. The fields are assumed to be axisymmetric. The assumption of symmetry allows
the analytic equation for the field lines to be derived. In reality, stellar magnetic fields can be highly complex, with many high order field components.
(See Ref.~\onlinecite{don09} for a review of recent results on stellar magnetic topologies.) A different approach is to model stellar magnetic
fields numerically via extrapolation from observationally derived magnetic maps.\cite{jar02,gre06,gre08} Both the numerical approach and the analytic 
work in this paper assume that the stellar magnetic fields are current free. As discussed in Sec.~\ref{comps_part2}, the global field topologies obtained 
in potential field source surface models approximately match those obtained from more computationally complex magnetohydrodynamic models.\cite{ril06,rom10}

In this paper I have assumed that the $\ell$th order multipole moment symmetry axis is aligned with the stellar rotation axis, and that both lie in the 
same stellar meridional plane (planes with $\phi={\rm constant}$; for example, the $x$-$z$ plane). However, Eqs~(\ref{final})--(\ref{final_source}) 
apply generally to tilted multipole symmetry axes (that is, large scale magnetospheres tilted arbitrarily in the polar and azimuthal directions with 
respect to the stellar rotation axis, assumed to be the $z$ axis) with appropriate coordinate and vector frame transformations.

The equation for calculating the co-latitude of the footpoints of the largest closed field line loop within each region of closed field lines can be used 
as a basis for calculating the amount of unsigned open (or closed) flux relative to the total flux (open plus closed) through the stellar surface. Such 
detailed calculations, from which models of stellar rotational evolution can be developed (for example, Refs.~\onlinecite{iva03} and \onlinecite{mat09}), 
are deferred to a future paper.

\appendix*
\section{\label{multi_basic}Basic properties of axial multipoles}
Consider a multipole of order $\ell$ with a single closed field line extending to a radius $r_{\max}$ above the stellar surface within each region of closed field lines. 
As $\theta$ increases from the rotation pole of the star ($\theta=0$) to the equatorial plane ($\theta=\pi/2$), the sign of $B_r$ and $B_\theta$ will vary according 
to whether $P_\ell(\cos \theta)$ and $P_{\ell1}(\cos{\theta})$ (respectively) are positive or negative. At the rotation pole of the star $P_\ell(\cos{\theta})=1$ 
and $P_{\ell1}(\cos{\theta})\propto \sin{\theta} = 0$. The field therefore only has a radial component, and $B_{\theta}=0$. As we move from the rotation pole 
toward the highest latitude closed field line footpoint, both $P_\ell(\cos{\theta})$ and $P_{\ell1}(\cos{\theta})$, and therefore $B_r$ and $B_\theta$, are positive. 
As $\theta$ is further increased, $P_\ell(\cos{\theta})$, and therefore $B_r$, changes sign every time we pass through the values of $\theta$ ($\Theta$ in our 
notation) where the closed field line loops reach their maximum radial extent (for odd $\ell$ multipoles there is a region of closed field lines that reach there maximum
radial extent in the stellar equatorial plane; for this region $\Theta=\pi/2$). 
$P_{\ell1}(\cos{\theta})$, and therefore $B_\theta$, changes sign at values of $\theta$ within the open field line regions that occur between the regions of 
closed field lines. Hence, for a particular closed field line with footpoints at $\theta_1$ and $\theta_2$ and which reaches its maximum radial extent 
$r_{\max}$ at $\Theta$, $P_{\ell1}(\cos{\theta})$ has the same sign at each point along the loop. For this reason the modulus signs in Eq.~(\ref{lhs_final}) 
can be removed.

\begin{figure}[h!]
\centering
\scalebox{0.70}{\includegraphics{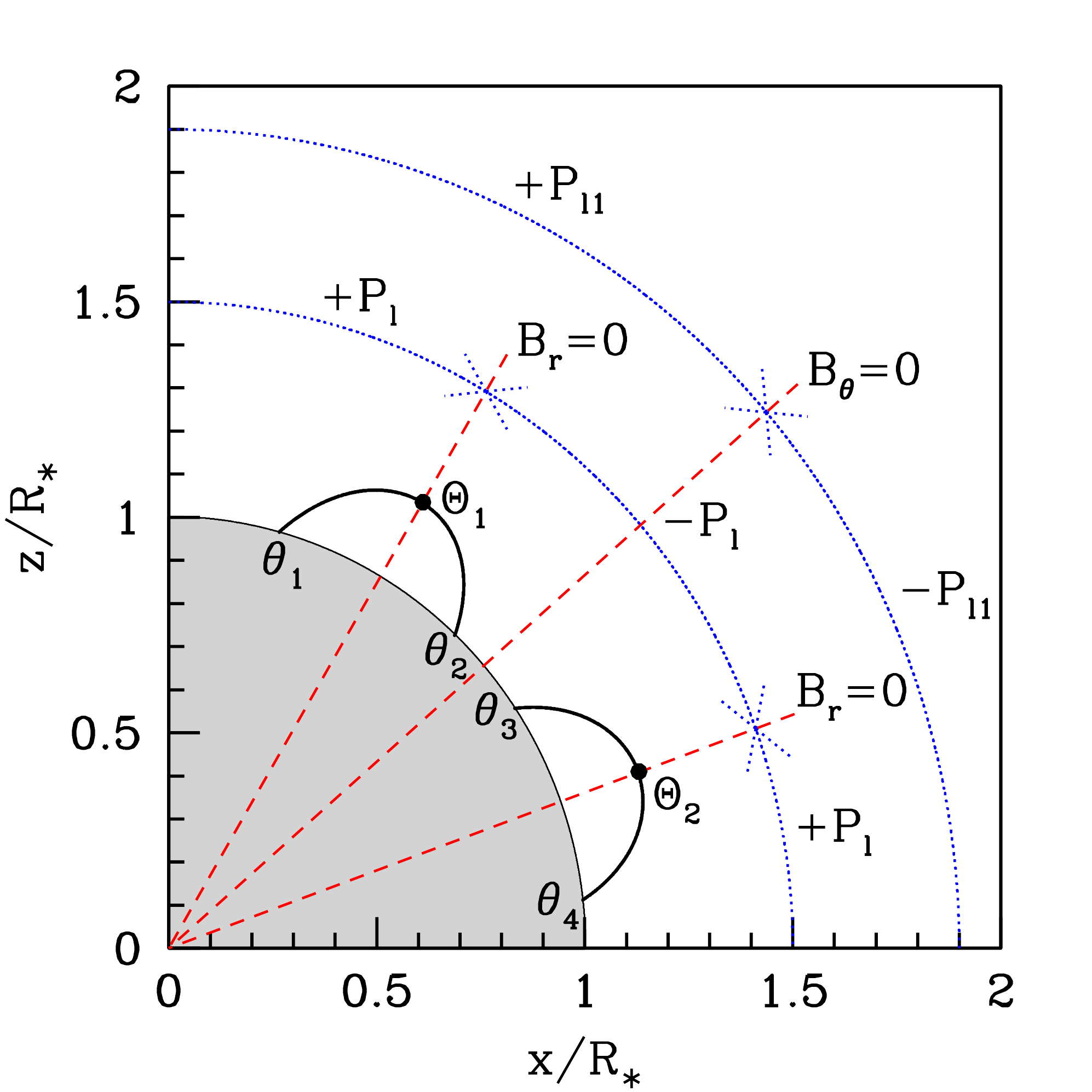}}
\caption{\label{fig2a}(color online) A star (shaded gray) with an axial hexadecapole magnetic field ($\ell=4$, $m=0$). $\Theta$ denotes 
where the closed field lines reach their maximum radial extent, and $\theta$ indicates the locations of the footpoints of the closed loops. 
For clarity only two closed field line loops within each closed field region are shown (solid lines), and no open field lines are shown. 
The field is rotationally symmetric about the $z$-axis and reflectionally symmetric about the $x$-axis, and therefore we need consider 
$\theta$ values only between zero and $\pi/2$. The field is purely radial at the stellar rotation pole, in the equatorial plane, and along a 
line between the closed field lines shown here (and by symmetry in the other three quadrants). Because $B_r\propto P_\ell(\cos{\theta})$ 
and $B_{\theta}\propto P_{\ell1}(\cos{\theta})$, the values of $\theta$ where $B_r$ and $B_{\theta}$ are zero correspond to the roots of 
the particular Legendre polynomial and associated Legendre function, respectively.}
\end{figure}

These ideas are best illustrated by considering an example of a particular multipole. Figure~\ref{fig2a} shows a 
hexadecapole ($\ell=4$) magnetic field. For clarity, only one loop within each region of 
closed field lines is drawn. The high latitude closed loop is assumed to have footpoints at $\theta_1$ and $\theta_2$, and the lower
latitude closed loop has footpoints at $\theta_3$ and $\theta_4$. The closed field lines in the upper region reach their maximum 
radial extent at $\Theta_1$, and the closed field lines in the lower region do so at $\Theta_2$. The closed loops are not symmetric 
about the $B_r=0$ lines because the Legendre polynomials are not symmetric about their roots. This property
means that the angular difference between each footpoint and where the particular region of closed field lines reach their 
maximum radial extent is not the same; for the particular hexadecapole field line $\Theta_1 - \theta_1 \ne \theta_2 - \Theta_1$. 
The exception to this is the equatorial region of closed loops for odd $\ell$ multipoles, which are symmetric about the stellar equatorial
plane.

\begin{acknowledgments}
This work was supported by the Science and Technology Facilities Council grant number ST/G006261/1. I thank the two referees to whom this
manuscript was sent simultaneously for 
constructive comments, and Prof. C.\ G.\ Gray (University of Guelph) for insightful comments and a detailed reading of the manuscript. 
\end{acknowledgments}

\section*{Figure captions}

\begin{thebibliography}{5}

\bibitem{jea27}J. H. Jeans,
\textsl{The Mathematical Theory of Electricity and Magnetism}, 5th ed. (Cambridge University Press, Cambridge, 1925).

\bibitem{ble76}B. I. Bleaney and B. Bleaney,
\textsl{Electricity and Magnetism}, 3rd ed. (Oxford University Press, Oxford, 1989), Vol. 1. 

\bibitem{hea95}M. A. Heald and J. B. Marion,
\textsl{Classical Electromagnetic Radiation}, 3rd ed. (Saunders College Publishing, Fort Worth, 1995). 

\bibitem{wyl99}H. W. Wyld,
\textsl{Mathematical Methods for Physics}, 2nd ed. (Benjamin/Cummings, Reading MA, 1999). 

\bibitem{gra84}C. G. Gray and K. E. Gubbins,
\textsl{Theory of Molecular Fluids} 
(Oxford University Press, Oxford, 1984), Vol. 1.

\bibitem{alt69}M. D. Altschuler and G. Newkirk Jr.,
``Magnetic fields and the structure of the solar corona I: Methods of calculating coronal fields,''
Sol. Phys. {\bf 9}, 131--149 (1969).

\bibitem{sch69}K. H. Schatten, J. M. Wilcox, and N. F. Ness,
``A model of interplanetary and coronal magnetic fields,''
Sol. Phys. {\bf 6}, 442--455 (1969).

\bibitem{jar02}M. Jardine, A. Collier Cameron, and J.-F. Donati, 
``The global magnetic topology of AB Doradus,''
Mon. Not. R. Astron. Soc. {\bf 333}, 339--346 (2002).

\bibitem{gre06}S. G. Gregory, M. Jardine, I. Simpson, and J.-F. Donati, 
``Mass accretion on to T Tauri stars,''
Mon. Not. R. Astron. Soc. {\bf 371}, 999--1013 (2006).

\bibitem{types}The spectral type of a star is a classification based on the presence and strength of certain 
spectral lines in the stellar spectra.

\bibitem{gre08}S. G. Gregory, S. P. Matt, J.-F. Donati, and M. Jardine,
``The non-dipolar magnetic fields of accreting T Tauri stars,'' 
Mon. R. Astron. Soc. {\bf 389}, 1839--1850 (2008).

\bibitem{don09}J.-F. Donati and J. D. Landstreet,
``Magnetic fields of non-degenerate stars,''
Annu. Rev. Astron. Astrophys. {\bf 47}, 333--370 (2009).

\bibitem{don06science}J.-F. Donati, T. Forveille, A. Collier Cameron, J. R. Barnes, X. Delfosse, M. M. Jardine, and J. A. Valenti,
``The large-scale axisymmetric magnetic topology of a very-low-mass fully convective star,''
Science {\bf 311}, 633--635 (2006).

\bibitem{don06}J.-F. Donati, I. D. Howarth, M. M. Jardine, P. Petit, C. Catala, J. D. Landstreet, J.-C. Bouret, E. Alecian, J. R. Barnes, T. Forveille, F. Paletou, and N. Manset,
``The surprising magnetic topology of $\tau$ Sco: fossil remnant or dynamo output?''
Mon. Not. R. Astron. Soc. {\bf 370}, 629--644 (2006).

\bibitem{sno94}T. P. Snow, H. J. G. L. M. Lamers, D. M. Lindholm, and A. P. Odell,
``An atlas of ultraviolet P Cygni profiles,''
Astrophys. J. Supplement Series {\bf 95}, 163--299 (1994).

\bibitem{don07}J.-F. Donati, M. M. Jardine, S. G. Gregory, P. Petit, J. Bouvier, C. Dougados, F. M{\'e}nard, A. C. Cameron, T. J. Harries, S. V. Jeffers, and F. Paletou, 
``Magnetic fields and accretion flows on the classical T Tauri star V2129 Oph,''
Mon. Not. R. Astron. Soc. {\bf 380}, 1297--1312 (2007).

\bibitem{gre10}S. G. Gregory, M. Jardine, C. G. Gray, and J.-F. Donati,
``The magnetic fields of forming solar-like stars,"
Rep. Prog. Phys. {\bf 73} 126901 (2010).

\bibitem{gra09}C. G. Gray, G. Karl, and V. A. Novikov,
``Quadrupolar contact fields: Theory and applications,'' Am. J. Phys. {\bf 77}, 807--817 (2009).

\bibitem{kir86}L. Kirkup,
``Magnetic field line simulation using a microcomputer,''
Phys. Educ. {\bf 21}, 107--110 (1986).

\bibitem{kno00}H. E. Knoepfel,
\textsl{Magnetic fields: A Comprehensive Theoretical Treatise for Practical Use} (John John Wiley \& Sons, New York, 2000).

\bibitem{smy50}W. R. Smythe,
\textsl{Static and Dynamic Electricity}, 2nd ed. (McGraw-Hill, New York, 1950).

\bibitem{kir85}L. Kirkup,
``Computer simulation of electric field lines,''
Phys. Educ. {\bf 20}, 142--145 (1985).

\bibitem{kri85}L. Kristjansson,
``On the drawing of lines of force and equipotentials,''
Phys. Teach. {\bf 23}, 202--206 (1985).

\bibitem{wil87}D. M. Willis and L. R. Young,
``Equation for the field lines of an axisymmetric magnetic multipole,''
Geophys. J. R. Astr. Soc. {\bf 89}, 1011--1022 (1987).

\bibitem{jef88}B. Jeffreys,
``Derivations of the equation for the field lines of an axisymmetric multipole,''
Geophys. J. {\bf 92}, 355--356 (1988).

\bibitem{bac88}G. E. Backus,
``The field lines of an axisymmetric magnetic field,''
Geophys. J. {\bf 93}, 413--417 (1988).

\bibitem{wil88}D. M. Willis and A. R. Gardiner,
``Equations for the field lines of a sectorial magnetic multipole,''
Geophys. J. {\bf 95}, 625--632 (1988).

\bibitem{gra79}C. G. Gray,
``Magnetic multipole expansions using the scalar potential,''
Am. J. Phys. {\bf 47}, 457--459 (1979).

\bibitem{kre99}E. Kreyszig,
\textsl{Advanced Engineering Mathematics}, 8th ed.
(John Wiley \& Sons, New York, 1999).

\bibitem{win05}D. E. Winch, D. J. Ivers, J. P. R. Turner, and R. J. Stening,
``Geomagnetism and Schmidt quasi-normalization,''
Geophys. J. Int. {\bf 160}, 487--504 (2005).

\bibitem{jar08}M. M. Jardine, S. G. Gregory, and J.-F. Donati, ``Coronal structure of the classical T Tauri star V2129 Oph,'' 
Mon. Not. R. Astron. Soc. {\bf 386}, 688--696 (2008).

\bibitem{sch78}M. Schulz, E. N. Frazier, and D. J. Boucher Jr., 
``Coronal magnetic-field model with non-spherical source surface,'' 
Sol. Phys. {\bf 60}, 83--104 (1978).

\bibitem{ril06}P. Riley, J. A. Linker, Z. Miki{\'c}, R. Lionello, S. A. Ledvina, and J. G. Luhmann,
``A comparison between global solar magnetohydrodynamic and potential field source surface model results,''
Astrophys. J. {\bf 653}, 1510--1516 (2006).

\bibitem{rom10}M. M. Romanova, M. Long, F. K. Lamb, A. K. Kulkarni, and J.-F. Donati,
``Global 3D Simulations of disc accretion onto the classical T Tauri Star V2129 Oph,'' in press [astro-ph/0912.1681].

\bibitem{sch96}M. Schulz and M. C. McNab, 
``Source-surface modeling of planetary magnetospheres,''
J. Geophys. Res. {\bf 101}, 5095--5118 (1996).

\bibitem{duf68}W. J. Duffin,
\textsl{Advanced Electricity and Magnetism for Undergraduates} (McGraw-Hill, London, 1968).

\bibitem{par58}E. N. Parker,
``Dynamics of the interplanetary gas and magnetic fields,'' 
Astrophys. J. {\bf 128}, 664--677 (1958).

\bibitem{erd53}A. Erd{\'e}lyi, W. Magnus, F. Oberhettinger, and F. G. Tricomi,
\textsl{Higher Transcendental Functions} (McGraw-Hill, New York, 1953), Vol. 1.

\bibitem{tas97} J.-L. Tassoul, \textsl{Stellar Rotation} (Cambridge University Press, Cambridge, 2000).

\bibitem{gra10}C. G. Gray, G. Karl, and V. A. Novikov,
``Magnetic multipolar contact fields: The anapole and related moments,''
Am. J. Phys. {\bf 78}, 936--948 (2010).

\bibitem{iva03}N. Ivanova and R. E. Taam,
``Magnetic braking revisited,''
Astrophys. J. {\bf 599}, 516--521 (2003).

\bibitem{mat09}S. P. Matt and R. E. Pudritz,
``New calculations of stellar wind torques,''
in \textsl{Cool Stars, Stellar Systems and the Sun: Proceedings of the 15th Cambridge Workshop on Cool Stars, Stellar Systems and the Sun}, edited by
H. C. Stempels (AIP, Melville, NY, 2009), p. 369.

\bibitem{foot1}By convention geophysicists use the Schmidt partially normalized form of the associated 
Legendre functions.\cite{wil87,win05} With this convention Eqs.~(\ref{Br}), (\ref{Btheta}), (\ref{Brsource}), 
and (\ref{Bthetasource}), and ultimately the derived equation for the field lines of an axial multipole with a source surface, Eq.~(\ref{final}), would be slightly modified. 
The difference, would only be by numerical constants, and the expressions would retain their functional form. Because the astrophysics community does not have 
accepted definitions of the spherical harmonics and the related associated Legendre functions, and to ensure consistency with previous work\cite{gre10}, 
I do not consider Schmidt partial normalization in this paper.

\end{thebibliography}
\end{document}